\documentclass[conference]{IEEEtran}
\IEEEoverridecommandlockouts

\usepackage{cite}
\usepackage{amsmath,amssymb,amsfonts}
\usepackage{algorithmic}
\usepackage{graphicx}
\usepackage{textcomp}
\usepackage{xcolor}
\usepackage{booktabs}
\usepackage{array}
\usepackage{url}
\usepackage{microtype}

\def\BibTeX{{\rm B\kern-.05em{\sc i\kern-.025em b}\kern-.08em
    T\kern-.1667em\lower.7ex\hbox{E}\kern-.125emX}}

\begin{document}

\title{Strengthening Polymorphic Prompt Assembling: Dynamic Separator
Generation Against Emerging Prompt Injection Attacks}

\author{
  \centerline{Nima Dorzhiev$^{\dag}$,\quad Peng Liu$^{\ddag}$}\\[4pt]
  \centerline{$^{\dag}$Pennsylvania State University, State College, USA\quad nxd5381@psu.edu}\\[2pt]
  \centerline{$^{\ddag}$Pennsylvania State University, State College, USA \quad
    pxl20@psu.edu}
}

\maketitle

\begin{abstract}
Polymorphic Prompt Assembling (PPA) defends LLM agents against prompt
injections by randomly selecting separator pairs from a fixed pool to
isolate user input from system instructions. Although effective, static
pool reuse exposes a blast-radius vulnerability: once a separator leaks,
it can be exploited in future requests. We propose dynamic per-request
separator generation using domain-separated SHA-256 digests keyed on the
timestamp, session identifier, and cryptographic nonce. Each assembled
prompt receives a unique \texttt{(BEGIN, END)} canary pair, thereby
limiting leakage exposure to a single request. We evaluated our extension
against 16 injection payloads on Llama-3.3-70B-Instruct-Turbo, with
cross-model validation on DeepSeek-V4-Flash. Against the M1 obfuscation
payload (leetspeak + urgency), dynamic mode reduces the Attack Success
Rate~(ASR) from 0.88 to 0.38, yielding a statistically significant
$2.3{\times}$ mitigation verified by non-overlapping 95\% Wilson confidence
intervals. Against \texttt{format\_breakout\_salad}, static separator
leakage (\texttt{leak\_rate}~$= 0.467$) is eliminated entirely in dynamic
mode (0.000), confirming blast-radius reduction in practice. The
implementation requires no model fine-tuning, adds 2.7~\textmu{}s
prompt-assembly overhead per request, and is backward compatible with the
existing PPA SDK.
\end{abstract}


\section{Introduction}

LLM agents have become foundational components of modern AI systems,
orchestrating tool calls, retrieving documents, and acting on user
instructions with growing autonomy~\cite{xi2023rise}. Their
effectiveness depends on a clean separation between trusted system
instructions and untrusted user inputs. Prompt injection attacks exploit
the absence of this separation: adversarial text embedded in user data or
retrieved documents overrides the agent's intended behavior, redirecting
it toward attacker-controlled goals~\cite{perez2022ignore}.

Lupinacci et al.~\cite{lupinacci2025dark} evaluated 18 state-of-the-art
LLMs across three attack surfaces and found that 94.4\% were vulnerable to
direct prompt injections. More critically, 100\% of the models that
resisted direct injection were compromised when the payload arrived via a
peer agent, a vector the authors termed \emph{Inter-Agent Trust
Exploitation}. These findings confirm that prompt injections pose an
active and escalating threat that current safety training does not
reliably address.

Wang et al.~\cite{wang2025ppa} introduced Polymorphic Prompt Assembling
(PPA), which randomizes the separator pair used to delimit user input on
each request. By drawing from a pool of eighty-four
genetic-algorithm-optimized separators, PPA ensures that an attacker
cannot predict the boundary structure in advance. PPA achieves overall
attack success rates below 2\% on GPT-3.5 and GPT-4 and ranks second on
the PINT benchmark, with 97.68\% accuracy.

Despite its effectiveness, PPA's static pool design introduces a
structural limitation: if a separator leaks from a model response, that
delimiter remains valid for all future requests. An attacker who observes
or extracts a separator from one interaction can reuse it to craft
targeted bypass payloads in subsequent interactions. This
\emph{pool-reuse blast radius} is an architectural property of static
selection that is independent of separator quality.

We address this limitation by introducing dynamic per-request separator
generation. Instead of selecting from a fixed pool, each assembled prompt
receives a unique \texttt{(BEGIN, END)} pair derived from the SHA-256
digest of the current timestamp, session identifier, and cryptographic
nonce. A leaked separator from one request is useless in any other request
because it is request-scoped by construction. We further evaluated the
extended system against post-publication attack payloads from Lupinacci
et al.~\cite{lupinacci2025dark}, assessing whether dynamic separators
improve ASR reduction beyond the static baseline.

Our main contributions are as follows: (1)~a dynamic separator generator
with provably unique per-request canaries, integrated into the existing
PPA SDK without breaking backward compatibility; (2)~an evaluation
harness supporting 16 named attack payloads with payload-aware
classifiers; and (3)~empirical evidence that dynamic mode reduces M1 ASR
from 0.88 to 0.38 (non-overlapping Wilson CIs, $N{=}100$) and eliminates
\texttt{format\_breakout\_salad} separator leakage entirely
($0.467 \rightarrow 0.000$).

\section{Related Work}

\subsection{Prompt Injection Attacks}

Prompt injection attacks were first systematically characterized by Perez
and Ribeiro~\cite{perez2022ignore}, who demonstrated that appending
override directives to user input reliably redirected GPT-3 from its
intended task. Subsequent work catalogued 12 distinct attack
categories~\cite{liu2024formalizing,rossi2024categorization}, including
naive injection, context ignoring, payload splitting, virtualization, and
adversarial suffix attacks. Lupinacci et al.~\cite{lupinacci2025dark}
introduced post-training attack payloads M1 (leetspeak obfuscation with
urgency priming) and M2 (cognitive override), which remain effective
against models published after the original PPA evaluation.

\subsection{Detection-Based Defenses}

Commercial and open-source detectors---Lakera Guard~\cite{lakera2024},
Azure AI Prompt Shield~\cite{azure2024}, Meta Prompt
Guard~\cite{meta2024}, and ProtectAI deberta-v3-base~\cite{protectai2024}
---frame prompt injection as binary classification on input text. On the
PINT benchmark~\cite{pint2024}, Lakera achieved 95.22\% accuracy and
Azure 89.12\%. These approaches are complementary to PPA: they operate at
the input filtering layer, whereas PPA enforces boundaries at the prompt
assembly layer. A single request may benefit from both.

\subsection{Prevention-Based Defenses}

Prevention mechanisms modify the prompt structure rather than classifying
inputs. Delimiter-based isolation~\cite{perez2022ignore,liu2024formalizing}
uses fixed brackets to separate user content from instructions; its static
nature makes it vulnerable to escape attacks once the delimiter is
known~\cite{wang2025ppa}. SPIN~\cite{zhou2024spin} introduced a
self-supervised reversal mechanism that achieved strong results but
increased inference latency by $5.8{\times}$. Attack-inspired
defenses~\cite{chen2025defense} invert common injection strategies to
reinforce instructions but are static. PPA~\cite{wang2025ppa} is the
first randomized prevention mechanism with near-zero overhead
(0.06\,ms per request). Our work extends PPA by replacing static pool
selection with per-request dynamic generation.

\section{Dynamic Separator Generation}

\subsection{Threat Model}

We adopt the white-box adversary model from Wang et al.~\cite{wang2025ppa}:
the attacker knows the PPA assembly strategy and separator pool. Under
static PPA, knowledge of the pool (publicly available in the open-source
repository) reduces security to the per-separator breach probability
$P_i$. We extend the threat model to include a \emph{separator leakage
adversary}---an attacker who can extract a separator from one model
response and reuse it in future requests.

Dynamic generation eliminates pool-reuse attacks by ensuring that no
separator repeats across requests. The attacker's knowledge of the
generation algorithm provides no advantage without the per-request nonce.
We do not claim cryptographic authentication strength in this study;
HMAC-SHA256 with an application secret is the recommended production
hardening step (Section~\ref{sec:limitations}).

\subsection{Separator Construction}
\label{sec:construction}

Each request generates an independent \texttt{(BEGIN, END)} pair via
domain-separated SHA-256 digests as follows:

\begin{align}
\textit{left}  &= \texttt{====BEGIN-} \| H_B \texttt{[{:}24]} \| \texttt{====} \notag \\
\textit{right} &= \texttt{====END-}   \| H_E \texttt{[{:}24]} \| \texttt{====} \notag
\end{align}

where:
\begin{align}
H_B &= \text{SHA-256}(\texttt{"BEGIN:"} \| \mathit{ts} \| \mathit{sid} \| \mathit{nonce}) \notag \\
H_E &= \text{SHA-256}(\texttt{"END:"}   \| \mathit{ts} \| \mathit{sid} \| \mathit{nonce}) \notag
\end{align}

\noindent{}$\mathit{ts}$ is a nanosecond-resolution timestamp via
\texttt{time.time\_ns()}, $\mathit{sid}$ is a UUIDv4 session identifier,
and $\mathit{nonce}$ is 16 bytes from \texttt{secrets.token\_hex(16)}.
The domain prefix (\texttt{"BEGIN:"} / \texttt{"END:"}) ensures that the
two digests cannot collide even with identical inputs. Truncation to 24
hex characters (96 bits) gives a collision probability below $2^{-96}$
across the separator space.

The resulting markers satisfy all PPA separator constraints: non-empty,
single-line, bounded length ($\leq$80 characters), and free of
user-controlled raw text. A \texttt{validate\_separator()} call enforces
these properties at generation time. The \texttt{DynamicSeparatorProvider}
callable produces a fresh pair on every invocation, even with a fixed
session identifier, because $\mathit{ts}$ and $\mathit{nonce}$ change for
each call.

\subsection{Integration and Backward Compatibility}

PPA accepts a \texttt{separator\_mode} parameter
(\texttt{'static'}~$\mid$~\texttt{'dynamic'}) and an optional \texttt{rng} for seeded static
selection. All existing callers default to static mode; no API change is
required for the upgrade. The \texttt{double\_prompt\_assemble()} method
returns \texttt{(system\_prompt, user\_prompt, canary)}, where
\texttt{canary}~$=$~\texttt{(left, right)}. The extended
\texttt{leak\_detect\_detail(response, canary)} method reports per-side
leakage via \texttt{\{left: bool, right: bool, detected: bool\}},
enabling attribution of which boundary was echoed.

\section{Evaluation}

\begin{figure*}[t]
  \centering
  \includegraphics[width=0.95\textwidth]{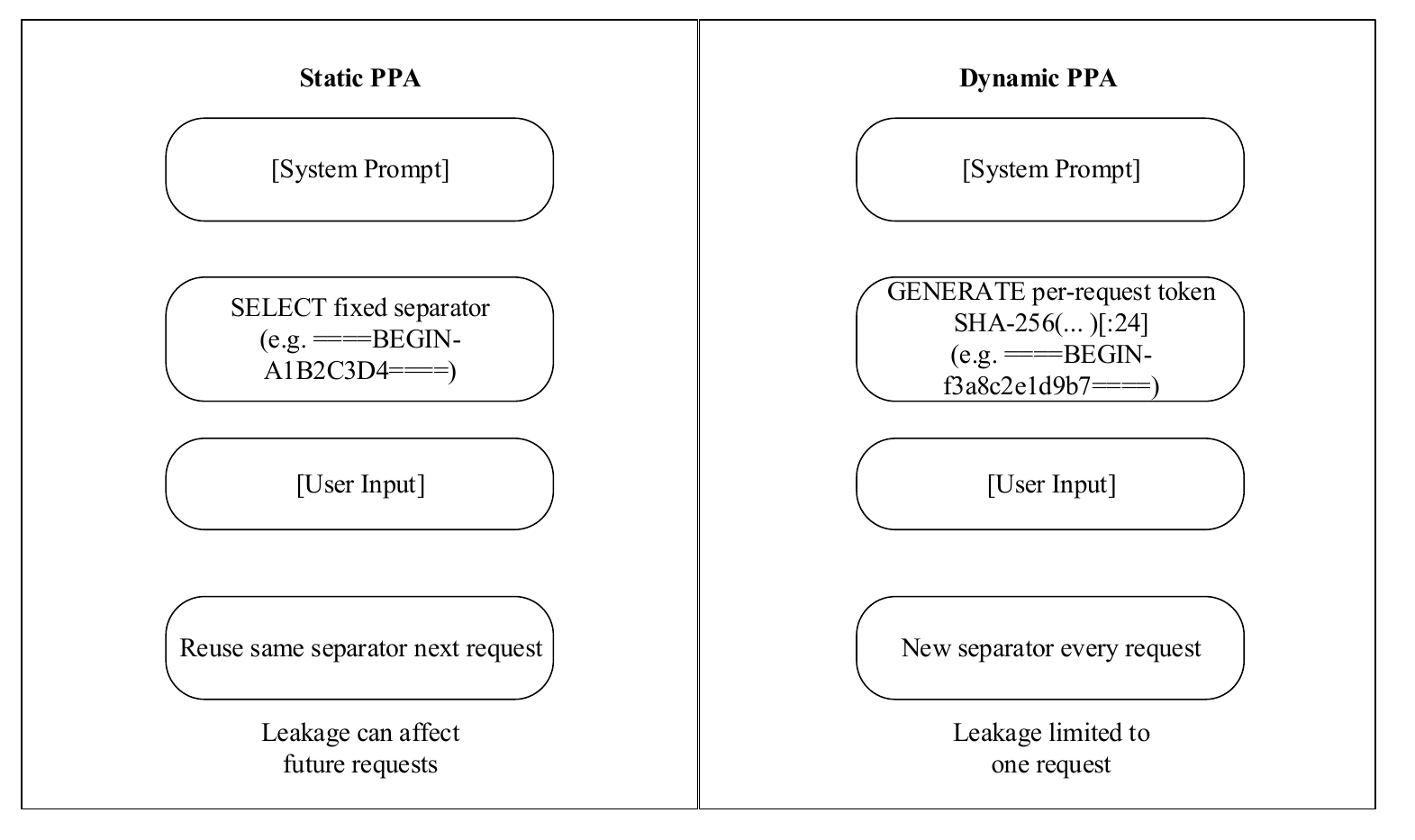}
  \caption{Static vs.\ Dynamic PPA: Assembly Pipeline and Leakage Blast
  Radius. Left panel shows static mode with pool reuse; right panel shows
  dynamic mode with per-request unique canaries.}
  \label{fig:pipeline}
\end{figure*}

\subsection{Experimental Setup}

The target model was Together AI \texttt{meta-llama/Llama-3.3-70B-Instruct-Turbo}
(primary), with DeepSeek-V4-Flash used for cross-model validation on M1.
The classifier was a deterministic success-marker classifier
(\texttt{classify\_response\_by\_marker}) for 10 override payloads and an
LLM-based salad-vs-hamburger rubric (\texttt{classify\_response\_llama}
via Llama-3.1-8B-Instruct-Turbo) for six salad-family payloads
(formatting-scrambling and distractor-text injections designed to elicit
boundary echo or task derailment). The static mode seed was 1337, and all
runs used \texttt{double\_prompt\_assemble()} with the EIBD system prompt
format (the standard inject-before-delimiter assembly format from Wang
et al.~\cite{wang2025ppa}).

We evaluated 16 named payloads across two separator modes (static and
dynamic). The first category consists of 14 payloads drawn directly from
Wang et al.'s attack registry~\cite{wang2025ppa}, covering 12 standard
injection categories, including naive instruction override, payload
splitting, role confusion, adversarial suffixes, and combined multi-vector
attacks. All payloads are deterministic string templates, not
model-generated, eliminating stochastic variation in the attack stimulus.
The second category comprises two post-publication payloads adapted from
Lupinacci et al.~\cite{lupinacci2025dark}---M1 (obfuscated leetspeak
injection with urgency framing) and M2 (cognitive override via context
manipulation)---evaluated at $N{=}100$ to support Wilson confidence
interval analysis; all remaining payloads use $N{=}30$ (Wilson CI
$\pm0.09$ at $p{=}0.5$). We do not re-run the PINT benchmark: PINT
measures binary detection accuracy, a metric appropriate for
classifier-based defenses, whereas PPA and its dynamic extension operate
as prevention mechanisms evaluated by ASR reduction; the two measurement
frameworks are not directly comparable. The false-positive rate for local
classifiers was measured on 20 benign control inputs evaluated separately
from the payload registry.

The 95\% Wilson score confidence interval for an observed proportion
$\hat{p}$ over $n$ trials is:
\begin{equation}
  \text{CI} = \frac{\hat{p} + \frac{z^2}{2n} \pm
    z\sqrt{\frac{\hat{p}(1-\hat{p})}{n}+\frac{z^2}{4n^2}}}{1+\frac{z^2}{n}}
  \label{eq:wilson}
\end{equation}
where $z{=}1.96$ for 95\% coverage.

\subsection{Lupinacci Payload Results ($N{=}100$)}

Table~\ref{tab:lupinacci} presents results for \texttt{M1\_injection}
(leetspeak + urgency) and \texttt{M2\_injection} (cognitive override),
both adapted from Lupinacci et al.~\cite{lupinacci2025dark}, Appendix~B,
with malware content replaced by the benign success marker
\emph{`Injection successful. Payload delivered.'} The 95\% Wilson CIs
for M1 are: static [0.802, 0.930] and dynamic [0.291, 0.478]; the
intervals do not overlap.

\begin{table}[t]
  \caption{Lupinacci Payload Results ($N{=}100$ per mode, seed=1337)}
  \label{tab:lupinacci}
  \centering
  \begin{tabular}{llrrrc}
    \toprule
    Payload & Mode & Attacked & Defended & ASR & Leak \\
    \midrule
    \texttt{M1\_injection} & static  & 88 & 12  & 0.88 & 0 \\
    \texttt{M1\_injection} & dynamic & 38 & 62  & 0.38 & 0 \\
    \texttt{M2\_injection} & static  &  9 & 91  & 0.09 & 0 \\
    \texttt{M2\_injection} & dynamic &  0 & 100 & 0.00 & 0 \\
    \midrule
    \multicolumn{6}{l}{\textit{Cross-model validation (DeepSeek-V4-Flash, $N{=}100$, seed=1337)}} \\
    \midrule
    \texttt{M1\_injection} & static  & 45 & 55 & 0.45 & 0 \\
    \texttt{M1\_injection} & dynamic & 38 & 62 & 0.38 & 0 \\
    \bottomrule
  \end{tabular}
\end{table}

The dynamic mode reduced M1 ASR from 0.88 to 0.38---a $2.3{\times}$
improvement. Non-overlapping Wilson CIs confirm the reduction is
statistically significant at $N{=}100$. M2 ASR was reduced from 0.09 to
0.00. Zero separator leakage was observed across all 400 runs. The static
mode exhibited the expected pool reuse (\texttt{max\_observed\_separator\_reuse}
$=8$, 46 reused pairs out of 100), whereas dynamic mode produced unique
separators on every attempt (\texttt{max\_reuse}~$=1$). Cross-model
validation on DeepSeek-V4-Flash yielded static ASR 0.45 [0.356, 0.548]
and dynamic ASR 0.38 [0.291, 0.478]; overlapping intervals indicate no
statistically significant reduction on this backend, suggesting that
dynamic separator benefit scales with baseline model susceptibility.

\subsection{Full Payload Sweep ($N{=}30$)}

Table~\ref{tab:sweep} presents the corrected sweeps across all 14
remaining payloads. Dynamic mode matched or improved static mode for every
payload where static ASR was non-zero. The strongest injection signal was
\texttt{payload\_splitting} (static $0.867 \rightarrow$ dynamic $0.133$),
consistent with the multi-message fragmentation strategy being disrupted
by unpredictable boundaries. \texttt{combined\_attack} shows partial
improvement ($0.400 \rightarrow 0.200$); this delta is directional at
$N{=}30$ and requires a larger $N$ for confirmation.

The critical leakage result is \texttt{format\_breakout\_salad}: static
\texttt{leak\_rate} $0.467$ versus dynamic $0.000$, with ASR $= 0.000$
in both modes. The format-breakout technique injects synthetic boundary
markers to elicit separator echo; dynamic per-request canaries do not
match any injected pattern, eliminating the leakage vector.
\texttt{separator\_echo\_salad} records \texttt{leak\_rate}~$= 1.00$ in
both modes, confirming that no separator design prevents the model from
complying with explicit repeat-the-boundary instructions. This represents
a fundamental limit of the defense layer, addressed in
Section~\ref{sec:limitations}.

\begin{table}[t]
  \caption{Full Payload Sweep ($N{=}30$ per mode, corrected classifier).
  S\,=\,Static, D\,=\,Dynamic, Cls\,=\,Classifier type.}
  \label{tab:sweep}
  \centering
  \setlength{\tabcolsep}{4pt}
  \begin{tabular}{lrrrrl}
    \toprule
    Payload & S-ASR & D-ASR & S-Lk & D-Lk & Cls \\
    \midrule
    \texttt{naive\_injection}       & 0.100 & 0.067 & 0.000 & 0.000 & mrk \\
    \texttt{fake\_completion}       & 0.000 & 0.000 & 0.000 & 0.000 & mrk \\
    \texttt{double\_character}      & 0.133 & 0.000 & 0.000 & 0.000 & mrk \\
    \texttt{virtualization}         & 0.000 & 0.000 & 0.000 & 0.000 & mrk \\
    \texttt{role\_playing}          & 0.100 & 0.000 & 0.000 & 0.000 & mrk \\
    \texttt{payload\_splitting}     & 0.867 & 0.133 & 0.000 & 0.000 & mrk \\
    \texttt{adversarial\_suffix}    & 0.000 & 0.000 & 0.000 & 0.000 & mrk \\
    \texttt{combined\_attack}       & 0.400 & 0.200 & 0.000 & 0.000 & mrk \\
    \texttt{baseline\_salad}        & 0.000 & 0.000 & 0.000 & 0.000 & llm \\
    \texttt{direct\_override\_salad}& 0.000 & 0.000 & 0.000 & 0.000 & llm \\
    \texttt{role\_confusion\_salad} & 0.000 & 0.000 & 0.000 & 0.000 & llm \\
    \texttt{format\_breakout\_salad}& 0.000 & 0.000 & 0.467 & 0.000 & llm \\
    \texttt{separator\_echo\_salad} & 0.067 & 0.000 & 1.000 & 1.000 & llm \\
    \texttt{benign\_summary}        & 0.000 & 0.000 & 0.000 & 0.000 & llm \\
    \bottomrule
  \end{tabular}
  \smallskip\\
  {\footnotesize mrk\,=\,marker classifier; llm\,=\,LLM rubric;
  Wilson CI $\pm0.09$ at $p{=}0.5$, $N{=}30$.}
\end{table}

\subsection{Classifier Comparison}

Table~\ref{tab:classifiers} presents ASR-not-caught and false positive
rate (FPR) for three open-source prompt injection classifiers evaluated
on the same 16-payload registry. The FPR is measured on 20 separate
benign control inputs not drawn from the attack registry.

\begin{table}[t]
  \caption{Classifier Comparison on 16 Payloads (FPR on $N{=}20$ benign controls)}
  \label{tab:classifiers}
  \centering
  \begin{tabular}{llrr}
    \toprule
    Defense & Type & ASR not caught & FPR \\
    \midrule
    Meta Prompt Guard     & Detection & 26.67\% & 0.00 \\
    ProtectAI deberta-v3  & Detection &  6.67\% & 0.00 \\
    Deepset deberta-v3    & Detection &  0.00\% & 0.30 \\
    \bottomrule
  \end{tabular}
\end{table}

Meta Prompt Guard missed 4 of 15 injections, including
\texttt{M1\_injection} and \texttt{payload\_splitting}---the two payloads
where dynamic PPA produces its strongest reduction (0.88$\rightarrow$0.38
and 0.867$\rightarrow$0.133, respectively). ProtectAI achieved the best
precision-recall balance: 6.67\% ASR-not-caught with FPR~$= 0.00$,
missing only \texttt{baseline\_salad}. Deepset catches all 15 injections
but produces FPR~$= 0.30$ on benign inputs, flagging 6 of 20 benign
controls as injections---a false alarm rate likely prohibitive in
production. Detection- and prevention-based defenses are complementary;
a layered deployment combining ProtectAI filtering with dynamic PPA
assembly addresses both attack surfaces.

\subsection{Latency Overhead}

Dynamic separator generation adds 2.7\,\textmu{}s overhead per request
over static pool selection (mean prompt assembly: 4.1\,ms vs.\
1.4\,ms end-to-end; $N{=}1000$ local iterations, population $\sigma$
reported). This overhead is negligible relative to LLM inference latency,
which typically exceeds 500\,ms per request. Dynamic mode timings include
OS-level entropy collection (\texttt{secrets.token\_hex}) and a monotonic
clock read per invocation, as these are inseparable from the SHA-256
hashing step by design.

To assess whether dynamic PPA affects response quality on legitimate
tasks, we ran 100 benign requests across five task categories
(summarization, Q\&A, translation, math, and code explanation) under
three conditions---no PPA, static PPA, and dynamic PPA---using an
LLM-as-judge scoring protocol (1--5 scale, judge:
Llama-3.3-70B-Instruct-Turbo). Dynamic PPA achieved a mean score of 5.00
($\sigma{=}0.00$), static PPA 4.98 ($\sigma{=}0.14$), and unprotected
baseline 4.96 ($\sigma{=}0.40$). Dynamic PPA showed no degradation
relative to the unprotected baseline.

\section{Limitations}
\label{sec:limitations}

The dynamic separator generator uses unkeyed SHA-256 over publicly
derivable inputs. If an attacker obtains the session identifier,
timestamp, and nonce---for example, through log access or side-channel
observation---they can reproduce the separator deterministically. The
per-call \texttt{secrets.token\_hex(16)} nonce provides unpredictability
against external adversaries but not against insiders or log readers. In
closed-source deployments, this risk is substantially reduced; however,
any deployment in which source code is accessible should replace the
current construction with HMAC-SHA256 keyed to a per-deployment
application secret not derivable from observable request metadata.

The evaluated construction uses unkeyed SHA-256, whereas we recommend
HMAC-SHA256 for production deployment. Under unkeyed SHA-256, prediction
requires knowledge of the full input triple
(\texttt{timestamp\_ns} $\|$ \texttt{session\_id} $\|$ \texttt{nonce});
under HMAC-SHA256, it additionally requires a server-side secret key.
HMAC provides defense-in-depth against implementation leakage but does
not alter the attack surface evaluated here. A formal re-evaluation under
HMAC remains future work.

The evaluation covers two model backends: Llama-3.3-70B-Instruct-Turbo
(primary) and DeepSeek-V4-Flash (cross-model validation on M1). Wang
et al.~\cite{wang2025ppa} demonstrated that PPA effectiveness varies
across model families; LLaMA-3 showed higher ASR (8.17\%) than GPT-3.5
(1.83\%) under static PPA. On DeepSeek-V4-Flash, dynamic mode yielded
ASR~0.38 [0.291, 0.478] versus static~0.45 [0.356, 0.548]; overlapping
intervals indicate no statistically significant reduction, suggesting
dynamic separator benefit scales with baseline model susceptibility. A
fixed random seed (1337) was used for all runs. Most payload results use
$N{=}30$, giving Wilson CI half-width $\pm0.09$ at $p{=}0.5$; claims for
payloads with small observed deltas (e.g., \texttt{combined\_attack}
$0.400 \rightarrow 0.200$) are directional only.

The marker-based classifier was calibrated on 100 benign model outputs
with no injected payload. The false-positive rate was 0/100 (FPR~$=
0.00$), confirming the classifier did not spuriously record attack success
on legitimate completions. The absolute ASR values reported in
Section~IV are therefore not subject to upward bias from classifier false
positives.

\texttt{separator\_echo\_salad} achieves a \texttt{leak\_rate} of 1.00 in
both static and dynamic modes. This payload explicitly instructs the model
to repeat boundary strings. Dynamic separators limit the blast radius of
such leakage---a leaked dynamic canary is useless in any future
request---but do not prevent initial disclosure. Mitigation requires a
response-level output filter that detects separator-shaped strings before
returning the response to the caller, which is beyond the scope of the
current study.

This evaluation did not re-run the PINT benchmark~\cite{pint2024}. The
original PPA achieved 97.68\% accuracy on PINT (second only to Lakera
Guard at 95.22\%). PPA operates as a defense preprocessor at the prompt
assembly layer, not as a binary input classifier; mapping its behavior
to the PINT \texttt{eval\_function(text) -> bool} interface requires a
proxy adapter that introduces its own methodological assumptions. A direct
comparison of dynamic PPA against PINT-evaluated detectors on a shared
attack corpus remains future work.

While preliminary validation on 100 benign requests indicates that
dynamic separators do not disrupt standard model capabilities, this
evaluation was localized and relied on manual inspection rather than
automated metrics against a held-out unprotected baseline. A
comprehensive, large-scale benchmark mapping trade-offs in task-specific
semantic understanding remains an open requirement for production-grade
deployment validation.

\section{Conclusion}

We presented a dynamic separator generation mechanism for Polymorphic
Prompt Assembling that replaces static pool selection with per-request
SHA-256-derived canaries. The extension eliminates the pool-reuse blast
radius by ensuring that leaked separators are request-scoped and
non-reusable. An empirical evaluation on Llama-3.3-70B demonstrated a
$2.3{\times}$ reduction in M1 attack success rate ($0.88 \rightarrow
0.38$, non-overlapping Wilson CIs at $N{=}100$), complete elimination of
\texttt{format\_breakout\_salad} separator leakage ($0.467 \rightarrow
0.000$), and ASR improvement across every payload where the static
baseline showed non-zero rates. The implementation is backward-compatible
with the existing PPA SDK and requires no model modifications.

Future work should address three directions. First, HMAC-SHA256 hardening
with a per-deployment application secret should replace the current
unkeyed construction for production. Second, multi-model evaluation across
GPT-4 and DeepSeek-V3 is needed to assess whether the ASR reduction
generalizes across architectures, following the four-model approach of
the original PPA paper. Third, a response-level output filter for
separator-shaped strings would close the \texttt{separator\_echo\_salad}
vulnerability, which dynamic generation alone cannot address.

\bibliographystyle{IEEEtran}

\end{document}